\documentclass{optica-article}
\journal{opticajournal}
\articletype{Research Article}
\usepackage[utf8]{inputenc}
\usepackage[english]{babel}
\usepackage{nicefrac}
\usepackage{graphicx}
\usepackage{wrapfig}
\usepackage{amsmath}
\usepackage{color}
\usepackage{ulem}
\usepackage{enumitem}
\usepackage{hyperref}
\usepackage{array}
\usepackage{notoccite}
\usepackage{listings}
\usepackage[T1]{fontenc}
\usepackage{bbold}
\usepackage{xspace}
\usepackage{textcomp}

\makeatletter
\let\ORIbbl@fixname\bbl@fixname
\def\bbl@fixname#1{
  \@ifundefined{languagealias@\expandafter\string#1}
    {\ORIbbl@fixname#1}
    {\edef\languagename{\@nameuse{languagealias@#1}}}
}
\newcommand{\definelanguagealias}[2]{
  \@namedef{languagealias@#1}{#2}
}
\makeatother

\definelanguagealias{en}{english}

\newcommand{\slevel}{$^2\mathrm{S}_{1/2}$\xspace}
\newcommand{\dlevel}{$^2\mathrm{D}_{5/2}$\xspace}

\newcommand{\be}{\begin{equation}}
\newcommand{\ee}{\end{equation}}

\definecolor{mypink1}{rgb}{0.858,0.188,0.478}
\definecolor{byzantium}{rgb}{0.44,0.16,0.39}
\definecolor{mygray}{gray}{0.5}
\definecolor{myblue}{rgb}{0.06,0.17,0.6}

\def\fm#1{\ifmmode #1 \else $#1$\fi}
\def\ket#1{{%
  \ifmmode |\,#1\,\rangle \else $|\,#1\,\rangle$\fi}}
\def\bra#1{{%
  \ifmmode \langle\,#1\,| \else $\langle\,#1\,|$\fi}}
\def\braket#1#2{{%
  \ifmmode \langle\,#1\,|\,#2\,\rangle \else $\langle\,#1\,|\,#2\,\rangle$\fi}}

\def\Ca{\fm{\mathrm{Ca}^{+}}\xspace}

\begin{document}
\title{A low phase noise cavity transmission self-injection locked laser system for atomic physics experiments}
\author{L. Krinner\authormark{1,2}, K. Dietze\authormark{2}, L. Pelzer\authormark{2}, N. Spethmann\authormark{2}, P. O. Schmidt\authormark{1,2,*}}

\address{\authormark{1}Leibniz Universit{\"a}t Hannover, Welfengarten 1, 30167 Hannover, Germany\\
\authormark{2}Physikalisch Technische Bundesanstalt, Bundesallee 100, 38116 Braunschweig, Germany}

\email{\authormark{*}piet.schmidt@quantummetrology.de}

\begin{abstract*}
Lasers with high spectral purity are indispensable for optical clocks and coherent manipulation of atomic and molecular qubits for applications such as quantum computing and quantum simulation. Stabilisation of the laser to a reference can provide a narrow linewidth and high spectral purity. However, widely-used diode lasers exhibit fast phase noise that prevents high fidelity qubit manipulation. Here we demonstrate a self-injection locked diode laser system utilizing a medium finesse cavity. The cavity not only provides a stable resonance frequency, but at the same time acts as a low-pass filter for phase noise beyond the cavity linewidth of around $100\,$kHz, resulting in low phase noise from dc to the injection lock limit.

We model the expected laser performance and benchmark it using a single trapped $^{40}$Ca$^{+}$-ion as a spectrum analyser. We show that the fast phase noise of the laser at relevant Fourier frequencies of $100\,$kHz to $>2\,$MHz is suppressed to a noise floor of between $-110\,$dBc/Hz and -120\,dBc/Hz, an improvement of 20 to $30\,$dB over state-of-the-art Pound-Drever-Hall-stabilized extended-cavity diode lasers. This strong suppression avoids incoherent (spurious) spin flips during manipulation of optical qubits and improves laser-driven gates in using diode lasers with applications in quantum logic spectroscopy, quantum simulation and quantum computation.
\end{abstract*}

\section{Introduction}
\label{section:introduction}

Semiconductor lasers have become an invaluable tool for the preparation, coherent manipulation, and spectroscopy of trapped atoms and ions \cite{wieman_using_1991, pavone_diode_1996, ricci_compact_1995, metcalf_laser_2007, stoehr_diode_2006, nasim_diode_2014, leibfried_quantum_2003, haffner_quantum_2008-1} with applications in quantum computing \cite{blatt_entangled_2008, pogorelov_compact_2021}, quantum simulation \cite{blatt_quantum_2012, scholl_quantum_2021, bloch_quantum_2018} as well as quantum sensing and metrology \cite{ludlow_optical_2015, pezze_quantum_2018, braun_quantum-enhanced_2018, degen_quantum_2017}. Their ever expanding range of directly accessible wavelengths along with their overall ease of use (setup, maintenance, spatial stability of output, compactness, energy efficiency and small footprint) have secured their use in atomic physics experiments. However, the spectral noise properties of the laser determine the quality of coherent control of qubits \cite{day_limits_2022}. Commercial distributed Bragg reflector (DBR) and distributed feedback (DFB) lasers exhibit linewidths of a few MHz \cite{nasim_diode_2014}, while extended cavity diode lasers (ECDLs) achieve linewidths of below $100\,$kHz \cite{saliba_linewidths_2009}.
Record laser linewidths below $10\,$mHz have been achieved by stabilizing the frequency of diode lasers to high-finesse cavities \cite{stoehr_diode_2006, matei20171_1, ito_stable_2017, hafner20158_1, keller2014simple_1, tarallo_high-stability_2010, dube_narrow_2009, alnis_subhertz_2008, ludlow_compact_2007}.

One of the major drawbacks of such semiconductor lasers is their significant noise at Fourier frequencies above a few $100\,$kHz \cite{zhang_quantum_1995}. Broadband amplified spontaneous emission (ASE) can be a limit in Doppler cooling \cite{schafer_optical_2015}, causes shifts in optical lattice clocks \cite{fasano_characterization_2021} by off-resonantly coupling ground and excited clock states to other levels and deteriorates the performance of Ramsey-Bordé atom interferometers \cite{nazarova_low-frequency-noise_2008}. Phase noise at frequencies that match motional frequencies of trapped atoms (100s of kHz up to several MHz) or nanoparticles \cite{meyer_resolved-sideband_2019} as well as mechanical resonances of optomechanical oscillators \cite{kippenberg_phase_2013} limit the sideband cooling performance.
The fidelity of laser-based entangling gates on trapped ions is similarly limited by the laser's noise spectrum incoherently coupling to the atomic qubit \cite{akerman_universal_2015, ballance_high-fidelity_2016, levine_high-fidelity_2018, day_limits_2022, nakav_effect_2023}. This is also the specific issue we will focus on here. 

One approach to overcome this limitation is to use solid-state ring-lasers (e.g. titanium-doped-sapphire, Ti:Sa) with intrinsically lower phase noise at these Fourier frequencies and practically no ASE \cite{fasano_characterization_2021} due to their lower round trip gain and longer excited state lifetime of the gain medium. These advantages typically come at a higher cost and footprint of the system as well as more frequent maintenance. A performance of similar quality can be achieved using fiber lasers, which, however, are only available at comparatively few wavelengths such as around 1.064\,$\mu$m and 1.5\,$\mu$m.

Cavity-stabilized diode lasers exhibit so-called "servo bumps" around the unity gain bandwidth of the servo loop.
In some cases feedback bandwidths of several megahertz have been achieved, thus pushing the servo bumps well above 1\,MHz \cite{kirilov_compact_2015, stoehr_diode_2006}, where they still limit applications sensitive in this frequency regime.

Linewidth narrowing down to a few kHz has been achieved by employing feedback from an optical cavity \cite{dahmani_frequency_1987, hollberg_modulatable_1988, breant_ultra-narrow_1989, simonsen_frequency_1993, labaziewicz_compact_2007, zhao_external_2011, jin_hertz-linewidth_2021}. Phase noise reduction of up to 20\,dB for Fourier frequencies up to 20\,MHz compared to a laser in Littrow configuration has been demonstrated by embedding a semiconductor gain medium in a ring cavity of 2\,m length \cite{celis_reducing_2019}. Optical feedback from an external fiber cavity to a DBR laser has shown to reduce the linewidth down to 300\,Hz with a noise suppression of 33\,dB for frequencies $>100$\,kHz \cite{lin_long-external-cavity_2012}.

High-finesse optical filter cavities with subsequent amplification of the transmitted laser radiation have been employed to further reduce high frequency noise \cite{akerman_universal_2015, fluhmann_encoding_2019, labaziewicz_compact_2007, hald_efficient_2005, nazarova_low-frequency-noise_2008}.  In an alternative approach the beat signal between the cavity-filtered light and the cavity-stabilized laser is employed to reduce the servo bumps through a high-speed electro-optical feedback loop \cite{li_active_2022}. Often, the cavity serves for frequency stabilization of the laser at the same time. Amplification of the usually weak ($<1$\,mW) transmitted beam is accomplished in a two-stage process by first injecting a laser diode, which then seeds a tapered amplifier.

In this work we demonstrate self-injection locking \cite{king_self-injection_2018,savchenkov_application_2020,liang2015compact,hao_narrow-linewidth_2021, zhao_high-finesse_2012, lewoczko-adamczyk_ultra-narrow_2015, kondratiev2017self} of a Fabry-Pérot diode to the low-pass filtered transmission of a linear medium-finesse cavity. This self-injection locked laser (SILL) inherits the frequency stability of the cavity for Fourier frequencies below the low-pass corner frequency of the optical cavity, while above it, the system exhibits the same phase noise as a laser injection-locked by the low-pass filtered output of the cavity. This way servo bumps in the critical frequency regime starting at around 10\,kHz are avoided and low phase noise is achieved all the way from dc to high Fourier frequencies. There, the performance is limited by the inherent phase noise of the source diode. The system can be conveniently upgraded from an existing semiconductor-laser, which is locked to a linear cavity. In section \ref{section:setup} we describe the laser and the apparatus, in section \ref{section:PNmodel} we describe how to translate the ion-excitation directly into phase noise and in section \ref{section:results} we present the excitation spectra obtained by the laser along with the resulting phase noise estimates. 

\section{Experimental Setup}
\label{section:setup}

\subsection{Self-injection locked laser system}

\begin{figure}[ht!]
\centering
    \includegraphics[width=1.0\columnwidth]{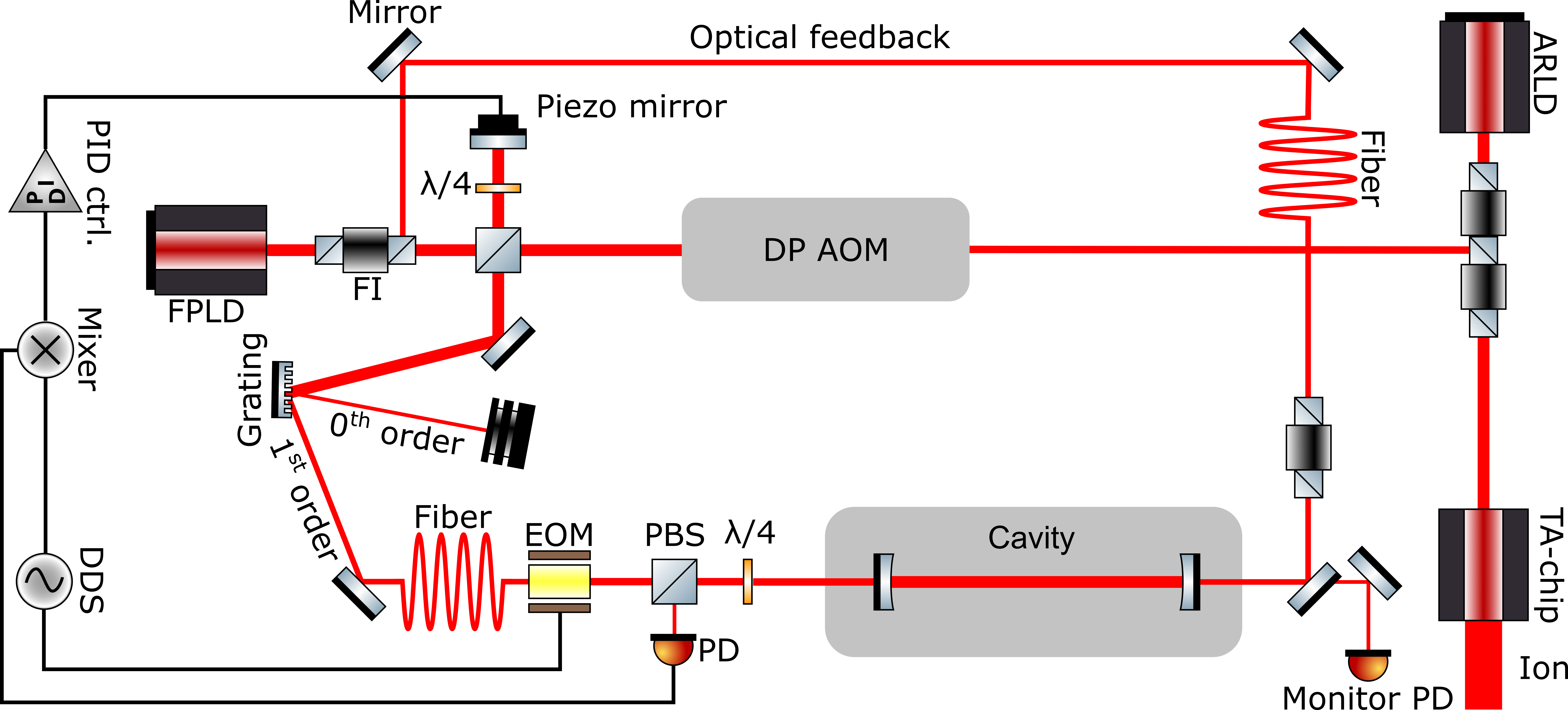}
    \caption{Laser setup for self-injection locking. The light for the laser is emitted from a Fabry-Perot laserdiode (FPLD), then sent over a path length modulation piezo mirror, a grating filter, a fiber, and an electro-optic modulator (EOM) for Pound-Drever-Hall (PDH) locking  to the medium finesse optical cavity. Its transmission is sent back to the FPLD and injects the laser via the rejection port of the laser's Faraday-isolator (FI). A fraction of the laser output is split off for seeding an antireflection coated laser diode (ARLD)/tapered amplifier (TA) combination after frequency shifting by a double-pass acousto-optic modulator (AOM). The unshifted part of the output is sent to a wavemeter for monitoring. Further abbreviations are PD: photodiode, PBS: polarizing beamsplitter, WP: waveplate. The component-library \cite{component_library} was used to create this graphic.}
    \label{fig:injection_setup}
\end{figure}

The injection setup (see Fig. \ref{fig:injection_setup}) consists of a Fabry-Pérot (Thorlabs HL7302MG) laser diode (FPLD) emitting at 729\,nm, which is seeded through the rejection port of a Faraday isolator by the beam transmitted through a medium finesse cavity. A piezo mirror between laser diode and cavity entrance is used to stabilize the path length to the cavity. The output of the laserdiode is spectrally filtered by a ruled grating (Thorlabs GR13-1208). The fiber after the grating to the cavity input is used as a spatial filter (not shown in Fig. \ref{fig:injection_setup}). 

Note that the ruled grating does not ensure or improve the operation of the self-injection lock, but rather helps to select the correct cavity mode for the laser to operate on. In principle any mode of the cavity inside the FPLD gain profile can begin to oscillate. The grating pre-selects modes within a $170~$GHz bandwidth (FWHM), corresponding to around 120 cavity modes resonant with the laser. This suppresses mode jumps to neighboring laser cavity modes, e.g., as a result of large external disturbances or the length stabilization piezo (see Fig.\ref{fig:injection_setup}) reaching the end of its travel and thus having to reset to the middle of its stroke.  

The employed cavity has a length of $10$\,cm and a finesse of $\mathcal{F}=11013(2)$, resulting in a full width half maximum linewidth of $\delta\nu\approx 140$\,kHz. It transmits approximately 10$\mu$W at an input power of 800$\mu$W. Just before the cavity we introduce a phase-modulation using an EOM (QUBIG, 12.7\,MHz). This allows us to implement a standard PDH locking scheme \cite{black_introduction_2001,pound_electronic_1946,drever_laser_1983} to stabilise the path length to the cavity for self-injection locking. 

\subsection{Ion trap apparatus}
We use the electronic excitation of a single trapped ion to determine an upper limit for laser phase noise as a function of laser detuning from the narrow \slevel-\dlevel resonance in \Ca{} \cite{nakav_effect_2023}. More detailed information about the spectral noise, especially at low Fourier frequencies, can be gained by employing a dynamical decoupling sequence \cite{bishof_optical_2013}. The ion trap setup is described in detail in \cite{hannig_development_2018, hannig_towards_2019}. In brief, we trap ions in a linear segmented Paul trap \cite{herschbach_linear_2012}, assembled from a stack of 4 laser-structured wafers made from Rogers$^{\mathrm{TM}}~$4350 material, with gold coating comprising the electrodes. 

The trapping frequencies along the three principal axes of motion in the trap are $(\omega_{x},\, \omega_{y},\, \omega_{z}) = 2\pi \times (1.34(5),\, 1.10(5),\, 1.26(2))~$MHz, utilizing a drive frequency of $\Omega_{r\!f} = 2\pi\times 33.0(1)~$MHz. Doppler-cooling followed by electromagnetically induced transparency (EIT) cooling \cite{morigi_ground_2000, scharnhorst_experimental_2018} is used to attain an average phonon occupation of  $(\bar{n}_x, \bar{n}_y, \bar{n}_z) \approx (0.5,\,0.5,\,0.5)$. 

A single ion is irradiated with light from the laser system either in regular ECDL/PDH configuration (see \cite{hannig_development_2018}) or in self-injection configuration for $200~\mu$s $\dots\,50~$ms. With these long probe times fast laser-noise can induce incoherent spin flips. The laser beam has an optical power of $\approx5~$mW at the ion. The excitation of the ion is read out via standard fluorescence detection using a detection time of $300~\mu$s. We scan the laser frequency from $-5\dots0.2~$MHz relative to the resonance of the 4\slevel$\,m_J = -1/2 \rightarrow$ 3\dlevel$\,m_J = -1/2$ transition in Ca$^+$ to determine the phase noise spectrum of the laser.
Typical on-resonance Rabi frequencies for this transition and our laser parameters range from $180\dots300~$kHz.

\section{Phase noise characterization}
\label{section:PNmodel}

In the following, we show how the atomic excitation can be converted to laser phase noise.
Following \cite{harty2013high, agarwal1976exact, agarwal1978quantum}, the time evolution of a two-level atom subject to a noisy field can be described as
\be\begin{split}
\frac{d\rho}{dt} &= -\frac{i}{2} \Omega [\hat{\sigma}_{x}, \rho]  -\frac{i}{2} \Delta [\hat{\sigma}_{z}, \rho] - \frac{\Omega^2}{8}\sigma_{y} \frac{N_0}{P_0} [\sigma_y,[\sigma_y,\rho]],
\label{eq:noise_master}
\end{split}\ee
where $\sigma_i$ are the familiar Pauli matrices, $\Omega$ denotes the coupling strength from ground to excited level, $\Delta$ is the detuning of the laser-radiation from resonance, and $N_0/P_0$ is the phase-noise power-spectral density when assuming a white-noise floor divided by the total power in the carrier. Since we assume small noise contributions, We will approximate the total power in the carrier by the total power in the laser.

Since the phase noise of the laser is not white but rather has a slowly varying frequency dependence, we will make the approximation that the noise is approximately white where it is near-resonant with the carrier, i.e. around the detuning $\Delta$ (only here we have slowly rotating terms that have a significant effect) by replacing $N_0/P_0$ with $10^{\frac{\mathcal{L}(\Delta)}{10}\mathrm{Hz}}/\mathrm{Hz}$ in Eq.~\ref{eq:noise_master}, where $\mathcal{L}(\Delta)$ is the single-sided phase noise in units of dBc/Hz. To a good approximation, a fully polarized input state ($p_g(0)=1$) yields the following solution under the time evolution of Eq.~\ref{eq:noise_master}: 
\be\begin{split}
p_{e}(t) &= \frac{\Omega^2}{\Omega^2+\Delta^2}\sin^2\left(\sqrt{\Omega^2+\Delta^2}\frac{t}{2}\right) \\
&+ 0.5\left(1-\frac{\Omega^2}{\Omega^2+\Delta^2}\right)\times\left(1-\exp\left(-\frac{\Omega^{2}}{2}\times10^{\frac{\mathcal{L}(\Delta)}{10\mathrm{dBc}}\mathrm{Hz}}\times \frac{t}{\mathrm{Hz}}\right)\right) ,
\end{split}
\ee
where the first part corresponds to the unitary dynamics, and the second part corresponds to the incoherent dynamics due to fast laser phase noise. The first part quickly dephases (due to magnetic and intensity fluctuations) to 0.5, allowing us to  extract the noise power spectral density (the Fourier frequency $2\pi\times f=\Delta$ is the detuning of the laser from atomic resonance):
\be\begin{split}
\mathcal{L}(f) = 10 \times \log_{10} \left ({  -2\ln\left(\frac{(1-2p_{e})\times((2\pi f)^2+\Omega^2)}{(2\pi f)^2}\right)}/{(t\Omega^{2}/\mathrm{Hz})
} \right )\mathrm{dBc/Hz}.
\label{eq:NPSD}
\end{split}\ee
We can see that in this approximation the noise close to the carrier has little observable effect, as it does not alter the steady state population ($\Omega>2\pi f$). Thus we limit the use of our model to at least one half of the Rabi-coupling strength away from resonance. At this point we estimate a minimum observable signal of 0.1 (quantum projection noise limited) on top of a background excitation of 40\,\% (from damped detuned carrier oscillation) using 100 repetitions of the experiment. This background reduces for larger detunings from the carrier.

\section{Experimental results}
\label{section:results}

\begin{figure}[ht!]
\centering
    \includegraphics[width=0.7\columnwidth]{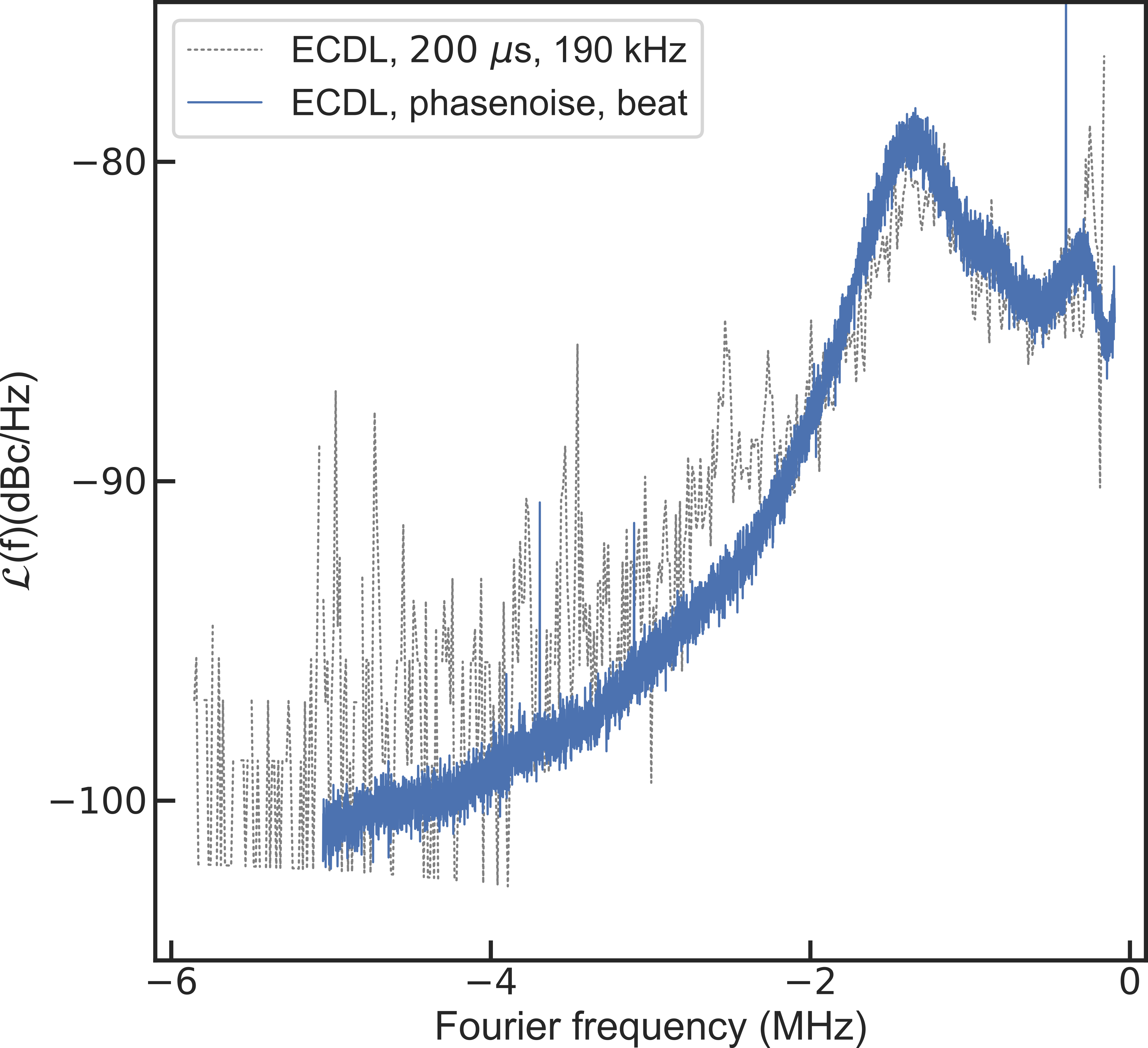}
    \caption{Comparison between ECDL phase noise derived from atomic excitation and a beat measurement. The ECDL is locked via the PDH technique to the medium finesse cavity in both cases. 
    Phase noise derived from atomic excitation is shown in dotted gray line, while the solid blue line is the beat signal between the output of the ECDL with the transmission of the cavity, amplified by injection locking another laser diode.
    }
    \label{fig:IonSA_ECDL}
\end{figure}

First, we verify the model for deriving the noise spectral density from atomic excitation developed in the previous section. For this we performed a beat measurement of the direct ECDL output and its low-pass filtered ($\approx 68$\,kHz corner frequency) transmission output from the cavity \cite{schmid_simple_2019} after amplification through injection locking of another laser diode, shown in Figure~\ref{fig:IonSA_ECDL}. 

\begin{figure}[ht!]
\centering
    \includegraphics[width=0.7\columnwidth]{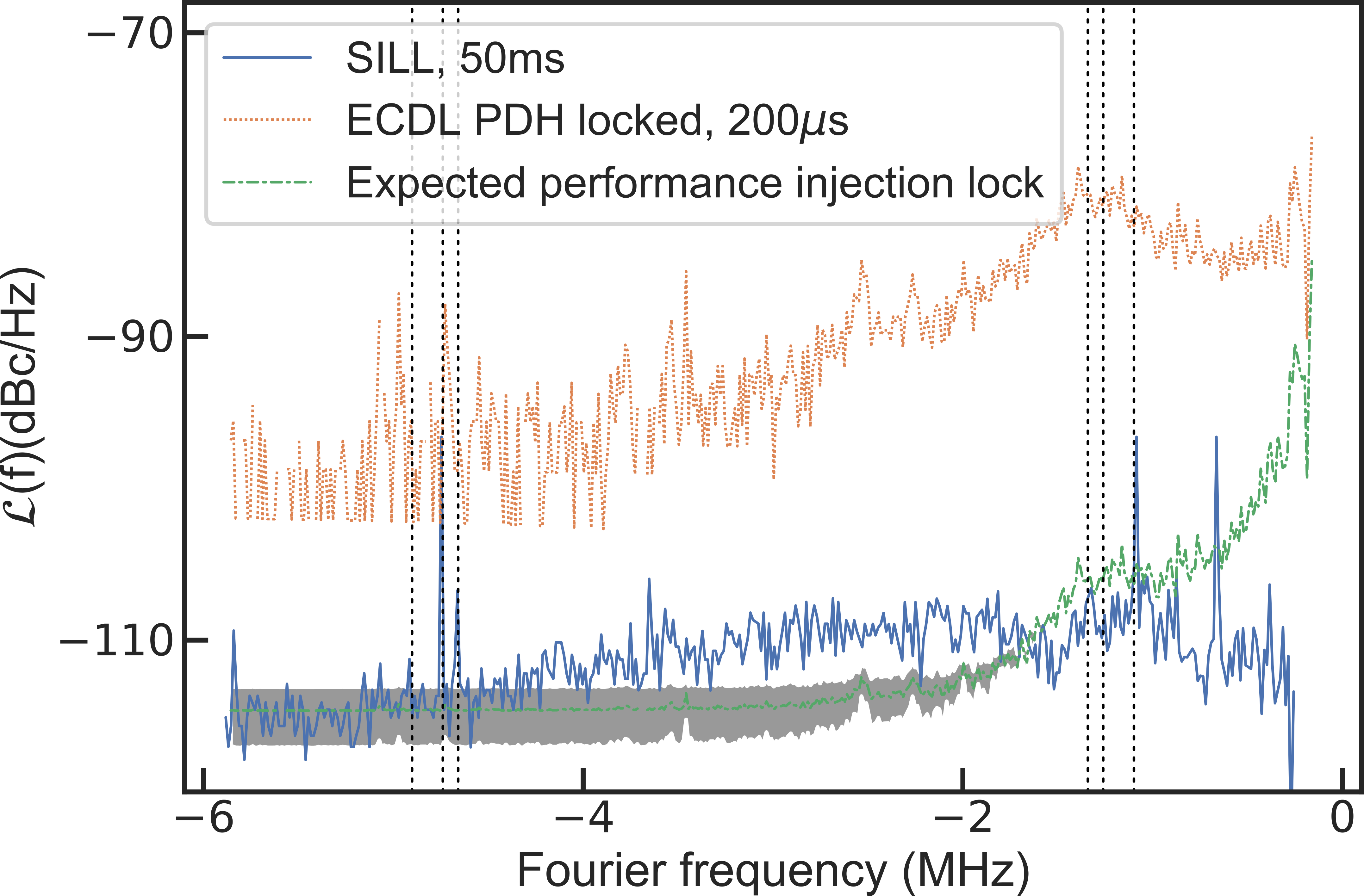}
    \caption{Phase noise calculated from incoherent excitation for the SILL (solid) and the ECDL (dotted). The dashed vertical lines indicate motional sideband peaks of the ion that are excited due to imperfect laser cooling and due to neighboring transitions. The green line (dash-dotted) shows the estimated performance of injection locking a slave laser diode to the light transmitted through the cavity. The white phase noise limit is estimated by the flat portion of the SILL spectrum between 5 MHz and 6 MHz. The shading indicates the uncertainty of the noise-floor estimation.}
    \label{fig:IonSA_SIL}
\end{figure}

Beyond Fourier frequencies of 200\,kHz, the phase noise of the transmitted light through the cavity is suppressed by more than 10\,dB and therefore has negligible effect on the beat measurement.
The injection locking is expected to limit noise performance at a broadband white-phase noise floor of approximately -115(2)dBc/Hz (see Fig. \ref{fig:IonSA_SIL}, which is extracted from the high-frequency performance of the SILL.
Figure~\ref{fig:IonSA_ECDL} shows good agreement between the inferred noise spectral density of the ECDL locked to the cavity from the beat signal and the result from ion excitation, confirming the noise excitation model. 

\begin{figure}[ht!]
\centering
    \includegraphics[width=0.7\columnwidth]{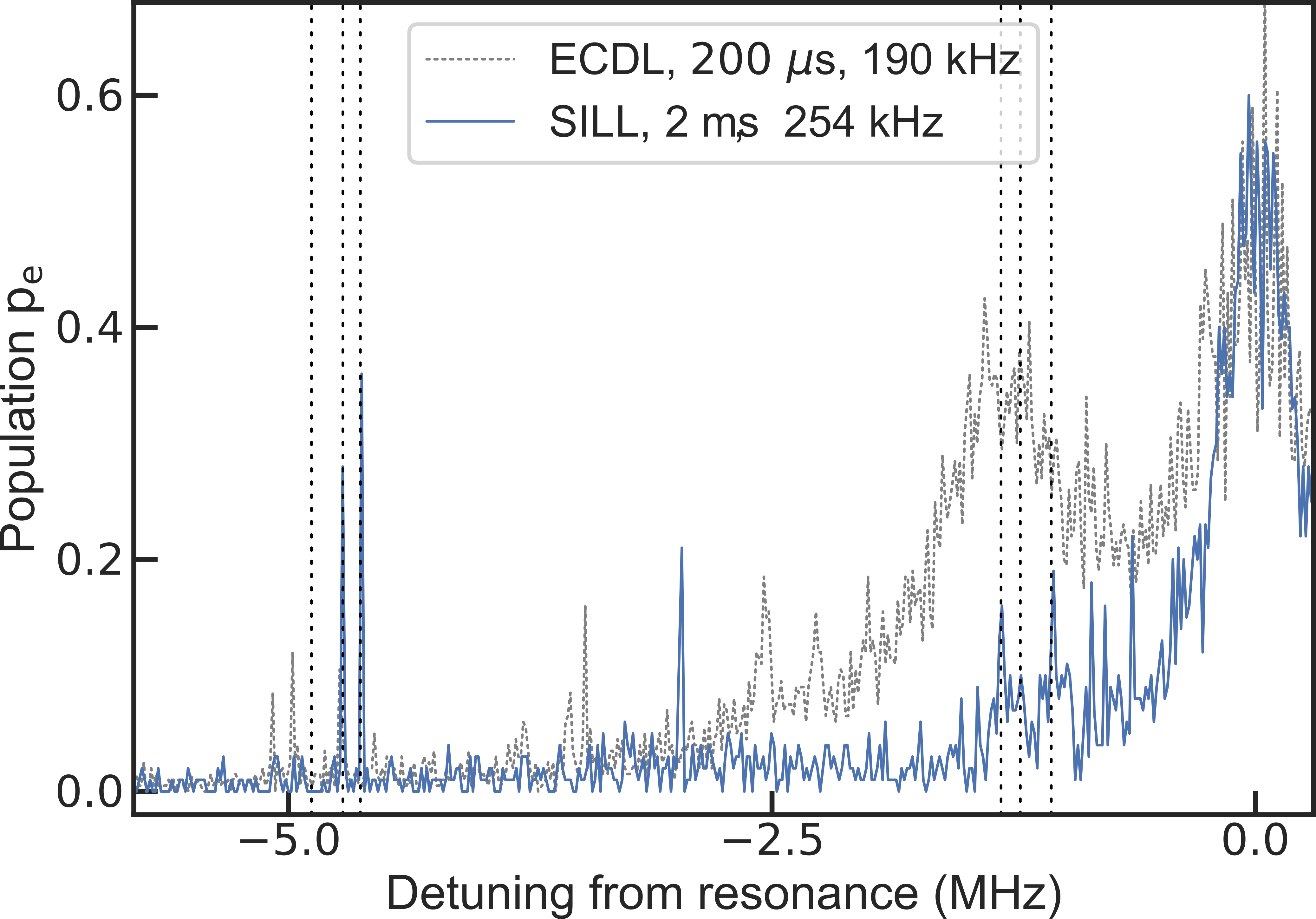}
    \caption{Comparative excitation spectra for ECDL (gray, dotted) and SILL (blue, solid). The ECDL is PDH-locked to a medium-finesse cavity, while for the SILL only the path length to the cavity is stabilized. The additional off-resonant excitation for the ECDL is clearly visible, where the shape of the noise is a consequence of the servo noise and locking bandwidth. The vertical dashed lines indicate motional sideband peaks of the ion that are accessible due to imperfect laser-cooling and due to neighboring transitions. }
    \label{fig:incoherent_exc}
\end{figure}

Next, the regular ECDL/TA laser is compared with the SILL setup by measuring the respective excitation spectrum from a single trapped ion. The result is shown in Fig.~\ref{fig:incoherent_exc}, showing a significant improvement of laser phase noise when employing the self-injection locking, especially considering the ten times longer effective probe time for the SILL compared to the ECDL/TA system and the larger Rabi frequency of $254\,$kHz for the SILL and $190\,$kHz for the ECDL/TA system.

In Fig.~\ref{fig:IonSA_SIL} we compare the noise spectral density of the ECDL with the SILL for different incoherent probe times. We observe a maximum noise suppression of the SILL compared to the ECDL of up to 30\,dB at a Fourier frequency of $1.5$\,MHz and of more than 20\,dB over the entire range from 0.1 to 6\,MHz. We have used probe times of $50\,$ms and $200\,\mu$s for the SILL and ECDL, respectively. The observed high-frequency noise floor is at the lower limit of what we estimate to be achievable with the Fabry-Perot laser diode we have used to build the SILL \cite{li_analysis_1989}.

\section{Discussion and summary}

In the following we discuss the experimental results in the context of theoretical models for optical injection locking and compare its performance to an injection-locked laser (ILL), seeded with the filtered light from the cavity.
The noise suppression factor $\eta=\frac{\Delta\omega_0}{\Delta\omega}$ is the ratio between the frequency noise of the free-running and self-injected laser. Its maximum value is given by \cite{li_analysis_1989, laurent_frequency_1989}:
\[\label{eq:eta}
\eta=1+\beta\sqrt{1+\alpha^2}\frac{Q_\mathrm{ext}}{Q_\mathrm{LD}},
\]
where $\beta=\frac{E_\mathrm{LD}}{E_\mathrm{inj}}$ is the feedback parameter (ratio between the electric field of the emitted and injected light), $\alpha$ is the linewidth enhancement factor of the laser diode, and $Q_\mathrm{ext}$ and $Q_\mathrm{LD}$ are the $Q$-factors of the external cavity and laser diode, respectively. In the SILL case, the feedback parameter $\eta$ is frequency dependent, since the cavity acts as a low-pass filter. 
The frequency-dependent noise floor of the SILL can be derived by considering the double-sided noise power spectral density $S(f)$ of an idealised injection lock of a perfect master oscillator into a slave laser diode \cite{spano_frequency_1986}:
\be\begin{split}
S_{f,\,\mathrm{slave}}(f) &\approx S_{f,\,\mathrm{master}}(f) + \frac{f^2}{\kappa^2} S_{f,\,\mathrm{free\,slave}}(f),
\label{eq:ILNF}
\end{split}\ee

where $f$ is the noise Fourier frequency, $\kappa^2 
\sim (I_{\mathrm{seed}}/I_{\mathrm{slave}})\Delta f_{\mathrm{FSR,\,LD}}^{2}$ is the optical feedback strength, and $f_{\mathrm{FSR,\,LD}}$ is the free spectral range of the laser diode. In the SILL setup, $\kappa$ is frequency dependent, since the seed intensity from the cavity depends on the detuning from resonance. Typical values for $\kappa$ on resonance of the SILL lie in the range of $(0.5 \dots 5$\,GHz, while relevant Fourier frequencies are on the order of hundreds of kilohertz to tens of megahertz, thus resulting in a noise suppression of the free running slave laser noise of up to $40\,$dB.

For detunings far beyond the cavity linewidth, $\kappa$ and thus the noise suppression drops to zero. According to Eq.~\ref{eq:ILNF}, in this region the achievable noise floor is thus dependent on the noise level inherent to the laser diode. The noise floor is a flat phase noise plateau, as the noise of Fabry-Perot laser diodes in this regime is to a good approximation white frequency noise.

A simplification of the SILL presented here is an injection-locked laser (ILL), which is seeded by the light filtered through an optical cavity without closing the optical loop to form a SILL \cite{akerman_universal_2015, fluhmann_encoding_2019, labaziewicz_compact_2007, hald_efficient_2005, nazarova_low-frequency-noise_2008}. Both systems will have a (close to) identical white phase noise floor beyond a certain Fourier frequency, dominated by the noise floor of the (slave) laser diode. 

For the SILL this limit is reached at the linewidth of the external cavity. For an ILL the limit depends on the details of the implementation, at which point the cavity-filtered servo bumps drop below the noise floor of the slave laser diode and will thus range between the cavity linewidth and the servo bump frequency.

Therefore, depending on the crucial frequency band for operation (in our case this band is $0.5~$MHz to $10~$MHz) and the used optical resonator, an ILL may have little operational difference to a SILL system. An advantage of the SILL is the large effective servo bandwidth from the optical feedback, which gives rise to improved noise suppression from dc up to Fourier frequencies of about the cavity linewidth. For very low Fourier frequencies the cavity instability itself as well as other noise sources will again overwhelm the achievable performance. 

A second interesting trade-off is the consideration, that the ultimate level of white phase noise is a function of power transmitted through the cavity (see Eq.~\ref{eq:NPSD}). The more power is transmitted through the cavity, the lower the white phase noise plateau, while the achievable linewidth in a laser lock to the cavity is inversely proportional to its finesse. Since finesse and transmission are typically anti-correlated, there is a trade-off between high frequency noise suppression and achievable laser linewidth. In our specific system, the linewidth of the SILL is further narrowed by transfer locking \cite{scharnhorst_high-bandwidth_2015} it to a narrow-linewidth reference laser \cite{matei20171_1}.

In summary, we have developed and characterized a SILL using the transmission of a medium-finesse cavity.  A theoretical framework for converting phase noise into incoherent electronic excitation has been developed that can be used to judge the suitability of a laser system for sideband operations. We find a suppression of laser phase noise by more than 30\,dB compared to a conventionally cavity-stabilized ECDL, approaching the theoretical limit set by the white frequency noise of the utilized laser diode at -115(2)\,dBc/Hz. Compared to an ILL, the noise suppression can be stronger for Fourier frequencies below the cavity linewidth due to the strong optical feedback. Noteworthy, the demonstrated level of phase noise enables two-qubit operations with an error of below $10^{-4}$ in trapped ion quantum computing using qubits based on a narrow optical transition \cite{nakav_effect_2023}.

\begin{backmatter}
\bmsection{Funding}
We gratefully acknowledge funding by the Deutsche Forschungsgemeinschaft (DFG, German Research Foundation), Project-ID 274200144 – SFB 1227 (DQ-\textit{mat}), Project-ID 434617780 – SFB 1464 (terraQ) and under Germany’s Excellence Strategy – EXC-2123 QuantumFrontiers – 390837967. This research project was ﬁnancally supported by the State of Lower Saxony, Hannover, Germany through Niedersächsisches Vorab. These projects 17FUN03 USOQS and 20FUN01 TSCAC have received funding from the EMPIR programme co-financed by the participating states and from the European Union’s 
Horizon 2020 research and innovation programme. This project has received funding from the European Research Council (ERC) under the European Union’s Horizon 2020 research and innovation programme (grant agreement no. 101019987).

\bmsection{Acknowledgments}
We thank Steven King and Fabian Wolf for careful review of the manuscript, and Klemens Hammerer and Adam Kaufman for helpful discussions on the manuscript. 

\bmsection{Data availability} Data underlying the results presented in this paper are not publicly available at this time but may be obtained from the authors upon reasonable request.
\end{backmatter}

\bibliography{EQM_Master, aux_bib} 
\newpage

\end{document}